\documentclass[
    ,final            
  ]
  {aipproc}

\layoutstyle{6x9}

\begin{document}

\title{Aspects of Cooling at the TRI$\mu$P Facility}

\classification{
                \texttt{11.30.Er, 32.80.Pj, 34.70.+e}}

\author{L. Willmann, G.P. Berg,  U. Dammalapati, S. De, P. Dendooven,
O. Dermois, K. Jungmann, A. Mol, C.J.G. Onderwater, A. Rogachevskiy, M.
Sohani, E. Traykov, and H.W. Wilschut}{
  address={Kernfysisch Versneller Instituut, Rijksuniversiteit Groningen,
  Zernikelaan 25, NL 9747 AA Groningen, The Netherlands} }

\begin{abstract}{
The Tri$\mu$P facility at KVI is dedicated to provide short lived
radioactive isotopes at low kinetic energies to users. It
comprised different cooling schemes for a variety of energy
ranges, from GeV down to the neV scale. The isotopes are produced
using beam of the AGOR cyclotron at KVI. They are separated from
the primary beam by a magnetic separator. A crucial part of such a
facility is the ability to stop and extract isotopes into a low
energy beamline which guides them to the experiment. In particular
we are investigating stopping in matter and buffer gases. After
the extraction the isotopes can be stored in neutral atoms or ion
traps for experiments. Our research includes precision studies of
nuclear $\beta$-decay through $\beta$-$\nu$ momentum correlations
as well as searches for permanent electric dipole moments in heavy
atomic systems like radium. Such experiments offer a large
potential for discovering new physics.}
\end{abstract}

\maketitle

Short lived radioactive isotopes are interesting for several
promising lines of research and currently a number of facilities
to provide the isotopes are planned or are being set up. The
research topics cover a wide range in atomic, nuclear and particle
physics \cite{schweikhard}. At the Kernfysisch Versneller
Instituut (KVI), Groningen, The Netherlands, we are commissioning
the TRI$\mu$P Facility (Trapped Radioactive Isotopes:
$\mu$icrolaboratories for fundamental Physics) \cite{trimp}, which
is open to outside users.

The physics interest of the TRI$\mu$P group are tests of discrete
fundamental symmetries, i.e. charge conjugation (C), space
inversion (P), and time reversal (T). In standard theory the
structure of weak interactions is V-A, which means that vector (V)
and axial vector (A) currents with opposite relative sign causing
a left handed structure and thus parity violation \cite{herczeg}.
Other possible currents like scalar, pseudo-scalar or tensor like
are signs for new physic. They can be tested by searching for
$\beta$-$\nu$ correlations in weak interactions \cite{betanu}. In
order to determine the neutrino momentum, the recoil to the
nucleus needs to be measured. Since the recoil energy is on the
order of 100 eV, a precision measurement of the recoil momentum
can only performed when the nuclei are suspended in a very shallow
potential, which can be provided by confining atoms by light
forces \cite{wilschut}. Particularly good candidates are
$^{20,21}$Na and $^{18,19}$Ne.

Another research direction is searching for permanent electric
dipole moments (edm), which violated C and P simultaneously
\cite{edmcp}. Any observation of an edm would be an indication of
physics beyond the standard model. Currently the most sensitive
experiment on a nuclear edm, was performed with $^{199}$Hg
\cite{fortson}, which gives a limit of 2.1$\times 10^{-28}$~ecm.
Recently Flambaum and collaborators pointed out that radium offers
a large sensitivity to edm's due to its nuclear as well as atomic
level structure \cite{flambaum}. Currently we are investigating
the feasibility of a search for an edm using radium. Both
experiments require to produce the isotopes and to store them
subsequently in a neutral atom traps. We will describe the setup
of the facility.


Isotopes of interest are produced in fragmentation or
fusion-evaporation reactions utilizing heavy ion beams from the
AGOR cyclotron on fixed targets, which are chosen for optimum
production rates. The production mechanism favors proton rich
isotopes. The goal is to provide a clean beam of the requested
isotopes with very low background radiations. This requires
dedicated separation and isotopes selective extraction stages. The
primary heavy ion beam is separated from the reactions product in
a magnetic device, which is designed to cover a wide range from
light to heavy isotopes. Through unique magnet design we achieved
a compact device. This magnetic separator has been successfully
commissioned in the fall of 2004 \cite{GPBerg}. After the
separator the reaction products are stopped in matter, which will
be discussed in below. The extraction time from the stopping
device sets the lower limit for the lifetime of the isotopes which
can be provided by the facility. We are aiming at times shorter
than 1~s. After the extraction from the stopping device the
isotopes are cooled and trapped in an radio frequency buncher
cooler (RFQ-cooler). From the buncher they are transported to the
experiment in an electrostatic beamline. After neutralisation the
atoms can be stored in neutral atom traps by laser cooling and
trapping methods.


A central role in the design of a facility for radioactive beams
takes the cooling from the energies of several MeV/u at the
productions to the eV range. For short lived isotopes the stopping
and extraction should be minimal. Reaction products stopped in
matter will be spread over a distance given by the range
straggling and initial momentum uncertainty. There are two
options.
\begin{itemize}

\item {Stopping ions in a buffer gas, preferentially helium. The
main common argument is that the ionization potential for helium
(24.5eV) is much larger than for any other elements. Thus the
neutralization of the incoming ion is energetically forbidden, at
least at low enough energies. The ionic isotope can be extracted
from the buffer gas by electrostatic guiding fields. A technical
aspect is that the helium gas has to be extremely clean. A main
question is the survival of ions in a buffer gas while they are
cooled by collisions especially at high rates of incoming ions
\cite{morrissey,huyse}. This approach is followed by groups at
several accelerators because it is less dependent on the specific
element.}

\item {Implanting in a solid which can be heated to a high
temperature at which the isotopes diffuse out of the material
\cite{Kirchner}. At high temperature the diffusion and effusion
time can be less than 1s if the stopping foils are sufficiently
thin ($\approx 1\mu$m). Such a thermoionizer could provide high
efficiencies of order 1 for alkaline and alkaline earth metals.}
\end{itemize}

\begin{figure}
  \includegraphics[height=.4\textheight,angle=-90]{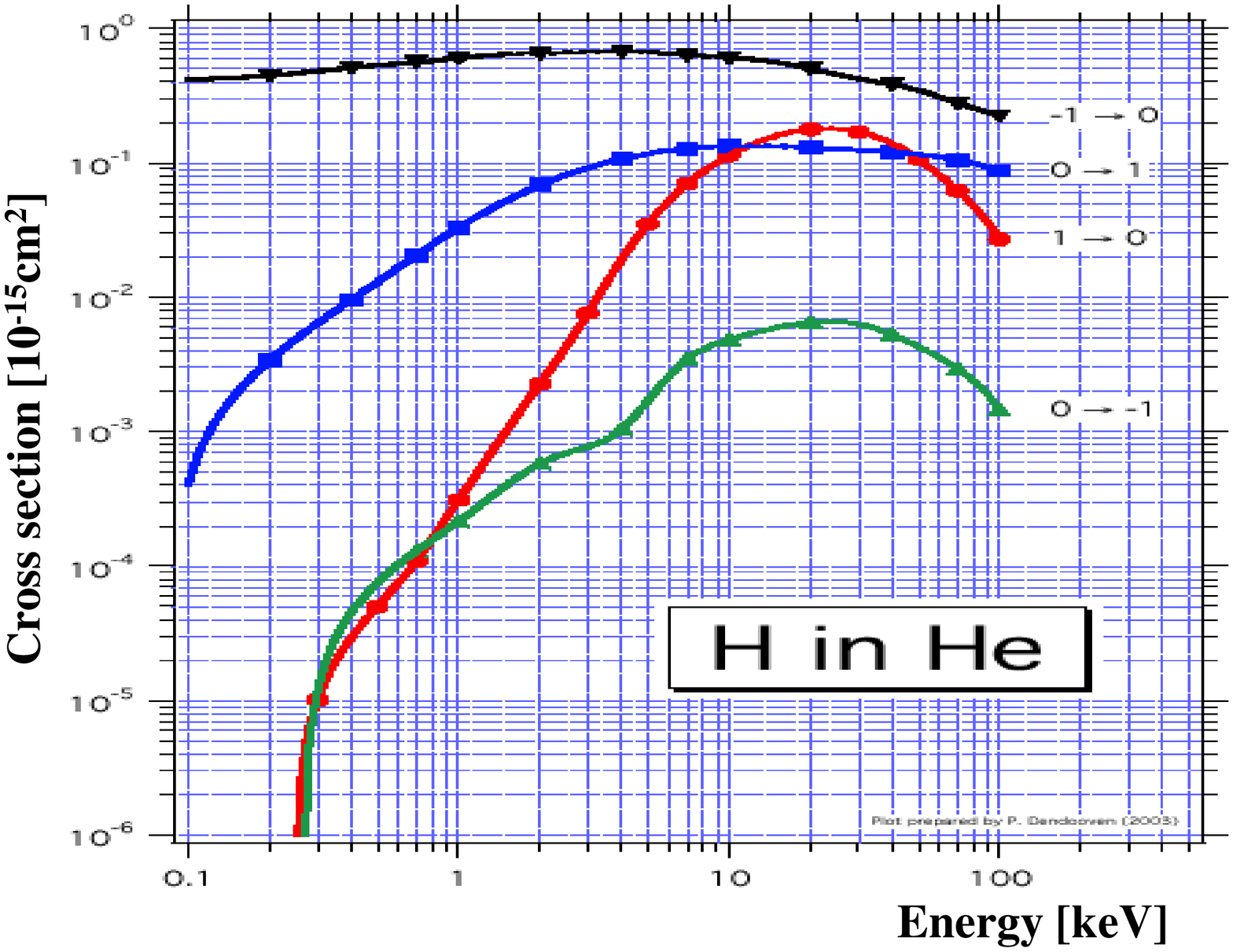}
  \caption{Charge exchange cross section for H on He \cite{hcrosssection}.}
    \label{hcross}
\end{figure}

We have investigated the possibility of using a gas cell. While an
ion traverses through the buffer gas it changes its charge state
very rapidly as it undergoes many neutralization-reionization
cycles. The neutralization has a kinematic cutoff because all
isotopes have a lower ionization potential then helium. The
neutralization cross section for a singly charged ion is maximal
around an energy of about 25~keV/nucleon (Fig. \ref{hcross}).
Below that energy the cross section drops rapidly. The fraction of
charged isotopes at thermal energies is expected to be determined
by the ratio of the ionization to neutralization cross section.
These charge exchange cross sections at low energies are important
input for the development of devices like radio frequency coolers
as well as gas catcher cells for stopping high energetic ions.

We have measured cross sections for neutralization at well as
stripping cross sections at energies below 1keV/u for different
buffer gases. The survival rate strongly depends on the
composition of the buffer gas. The preferred choices for a buffer
gas is highly pure inert gases where helium stands out because of
its high ionization potential of 24.5eV.

\begin{figure}
  \includegraphics[height=.5\textheight,angle=270]{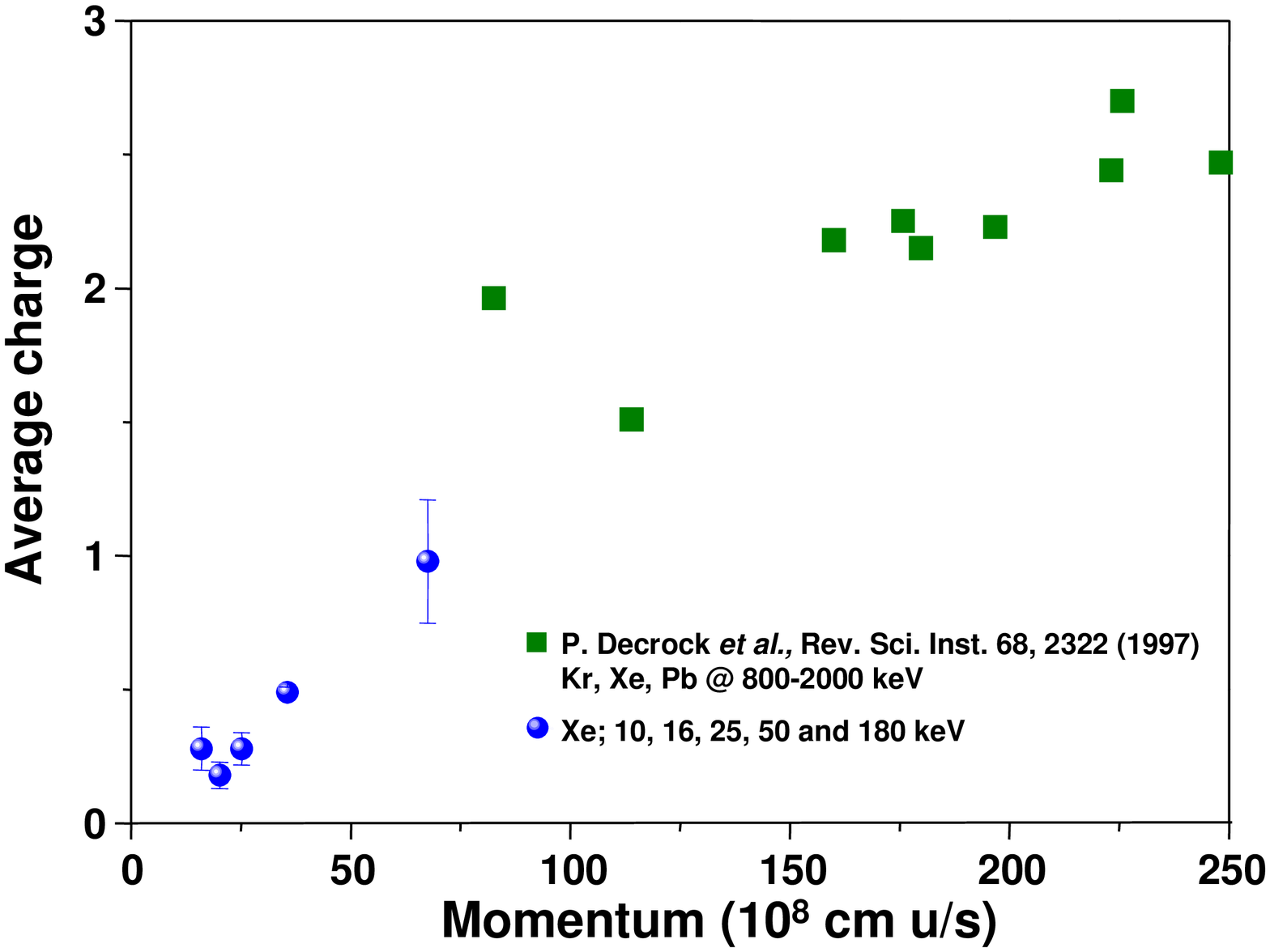}
  \caption{Average charge state of ions passing through helium at different
  momenta. The Xe data are from our measurements.}
  \label{figa}
\end{figure}

We measured the change of the charge state of multiply charged
ions after passing through a differentially pumped He gas target.
Typically less than 10 collisions are sufficient to reach an
equilibrium charge state distribution, while the energy loss is
small compared to the total energy. Because of the small number of
collisions such a setup is less sensitive to the purity of the gas
than in a measurement where we completely stop the ions. In
Fig.\ref{figa} our results for Xe on helium is plotted in addition
to data at higher energies and for different isotopes. We could
extend these measurements to lower momentum of 15$\times
10^8$cm~u/s, respectively 80 eV/u of energy. The average charge
state is decreasing with decreasing momentum of the ions.
Measurements at lower momenta were not possible in our apparatus
because of the increasing scattering angle.

For the TRI$\mu$P facility we are currently commissioning a
stopping device of the second type, since it is ideal for alkaline
and alkaline earth isotopes. We plan to stopping the ions in thin
tungsten foils, which can be heated to temperatures of 2500~K were
the expected diffusion times are less than 1~s for 1$\mu$m thick
foils.

\begin{figure}
  \includegraphics[height=.57\textheight,angle=270]{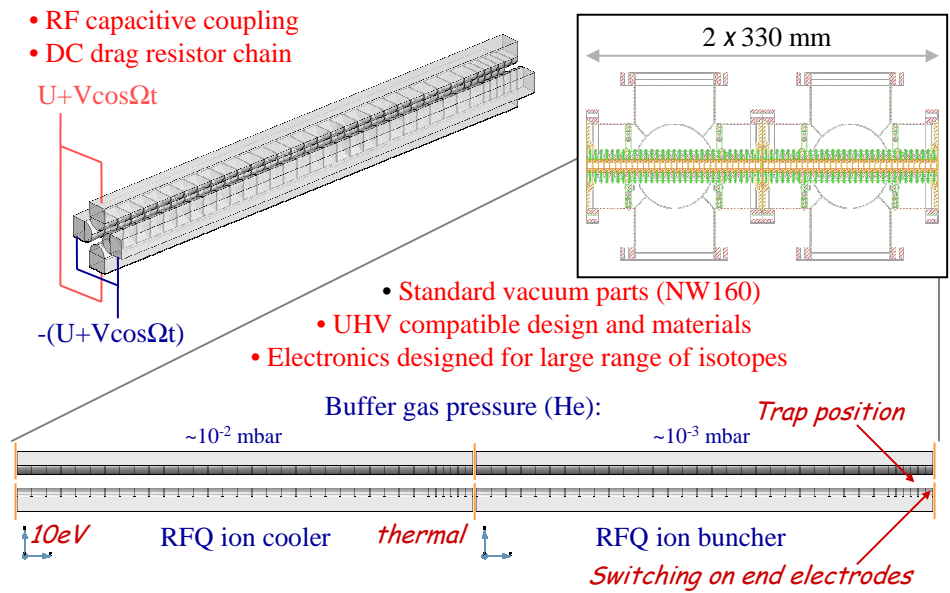}
\vspace{-1cm}
  \caption{The Radio Frequency Cooler Buncher for the TRI$\mu$P Facility.}
\label{rfqsetup}
\end{figure}

After the stopping device the isotopes are extracted as ions,
which allows for easy manipulation. The ions are guided by
electrostatic means into a gas filled radio-frequency buncher
cooler system. It consists out of two identical segmented rfq's of
330~mm length, which are separated by small apertures for
differential pumping purposes. The rods of the quadrupoles are
10~mm apart and are segmented in order to apply axial field
gradient. The segments are connected by a dc resistor chain, while
the rf is capacitively coupled to the segments. This reduces the
number of electrical feedthroughs significantly. We use
frequencies from 0.5-2~MHz and a voltage V$_{pp}$ of up to 200~V,
which is sufficient for the isotopes of interest. The device is
housed in standard UHV double crosses (Fig. \ref{rfqsetup}).

The system was commissioned using ions with energies of 10-60~eV.
In the first section they are slowed and transversely cooled by
collisions with He buffer gas of about 3$\times 10^{-2}$~mbar. A
small drag potential of 0.5~V/cm moves the ions along the axis.
The second rfq operates at a pressure ten times lower. The axial
potential has is shaped to allow for trapping near to the exit.
The potential depth is on the order of 5~V The ions can be ejected
into a pulsed drift tube accelerator by switching the last
electrode by several ten volt. Preliminary measurements indicate
that we find more than 60$\%$ of the ions entering the device are
transferred into the drift tube. The drift tube is pulsed and the
ion pulse is detected by a micro channel plate in the low energy
beamline. The pulse width is in the order of several hundredths of
ns in agreement with simulations. The ion pulse can then be
transported in an electrostatic beamline to the experiments.

The ultimate cooling of the radioactive isotopes is provided in
neutral atom traps. Here, atoms are well localized at typical
temperatures of the order of $\mu$K. Storage times in a MOT
depends on the particular atom and the background pressure and Na
trapping times of more than 100~s have been achieved. Recently the
first Na MOT for TRI$\mu$P has been brought into operation. An
advantage of optical trapping is that it allows to manipulate the
state of the systems. The limitation of laser cooling are that the
forces are rather small and atomic level scheme has to be
suitable. Thus we are developing new laser cooling schemes for
atoms like radium, extending the list of trapable atoms.

Research with trapped rare isotopes offer unique possibilities for
testing fundamental interactions in a complementary way to high
energy physics. Atomic physics techniques allow for precision
measurements which can test extensions to the standard model very
sensitively \cite{fortson,wieman}. The upcoming facilities at KVI
and other places are on their way to enable promising experimental
test in the near future.

\bibliographystyle{aipproc}   

\begin{thebibliography}{99}

\bibitem{schweikhard} Atomic Physics at Accelerators: Laser Spectroscopy
and Applications, L. Schweikhard and H.J. Kluge (eds.), Hyperfine
Interactions 127.

\bibitem{trimp} For more informations in the TRI$\mu$P Facility:
http://www.kvi.nl/~trimp/web/html/trimp.html.

\bibitem{herczeg}  P. Herczeg, Precision Tests of the Standard
Electroweak Model (World Scientific, Singapore, 1995).

\bibitem{betanu}  N. D. Scielzo et al., Phys. Rev. Lett. 93, 102501
(2004); A. Gorelov et al., Phys. Rev. Lett. 94, 142501 (2005).

\bibitem{wilschut} J.W. Turkstra, H.W. Wilschut, D. Meyer, R. Hoekstra, R.
Morgenstern, Hyperfine Interactions 127,  533-536 (2000).

\bibitem{edmcp} "CP Violation without Strangeness", I.B.
Krhiplovich, S.K. Lamoreaux, Springer, Berlin (1997).

\bibitem{fortson} M. V. Romalis et al. Phys. Rev. Lett. 86, 2505 (2001).

\bibitem{flambaum} V.V. Flambaum, Phys. Rev. A 60, R2611 (1999);
 V. A. Dzuba et al. Phys. Rev. A61 062509 (2000).

\bibitem{GPBerg}
G.P.A. Berg et al., accepted for publication Nucl. Inst. Meth. A,
xxx.lanl.gov:nucl-ex/0509013

\bibitem{morrissey} L. Weissman, P.A. Lofy, D. A. Davies,
D.J. Morrissey, P. Schury, S. Schwarz, and G. Bollen, Nucl. Phys.
A746c 655 (2004).

\bibitem{huyse}
M. Huyse, M. Facina, Yu.Kudryavtsev, P.Van Duppen, Nucl. Instr.
Meth. B187, 535 (2002).

\bibitem{Kirchner} R. Kirchner, Rev. Sci. Instrum. 67 928 (1996).

\bibitem{hcrosssection} Atomic Data for Fusion, Volume 1, C.F. Barnett editor,
ORNL 6086. and from R.Hoekstra, H.P. Summers and F.J. de Heer,
Nucl. Fusion Suppl. 3, 63 (1992).

\bibitem{wieman} C.S. Wood et al., Science 275, 1759 (1997).


\end{thebibliography}

\end{document}